# A symbolic calculus for a class of quantum computing circuits

F.Z. Hadjam, C. Moraga

This paper introduces a symbolic calculus to evaluate the output signals at the target line(s) of quantum computing subcircuits using controlled negations and controlled-Q gates, where Q represents the k-th root of [0 1; 1 0], the unitary matrix of NOT, and k is a power of two. The controlling signals are GF(2) expressions possibly including Boolean expressions. The method does not require operating with complex-valued matrices. The method may be used to verify the functionality and to check for possible minimization of a given quantum computing circuit using target lines. The method does not apply for a whole circuit if there are interactions among target lines. In this case the method applies for the independent subcircuits.

*Introduction:* Reversible switching theory represents a fast developing area, because of its possible applications in low power circuits, nano-electromechanical systems, superconducting interference devices, and because in the field of quantum computing, all basic operations must be reversible [1]. Much effort has been dedicated to develop methods to design reversible/quantum digital circuits (see e.g. [2] to [8]) focusing on benchmarks data bases (e.g. [9], [10]). One of the basic conditions to be fulfilled by a reversible/quantum circuits design system is to satisfy a given specification. A validation requires calculation of the outputs for specified, mostly all, inputs. A possibly semi-straightforward, formally correct method to do this, resembles the "bit-slices" architectures of the decade of the former 70s. A bit-slice comprises independent gates at the same depth in a circuit. The transfer matrix of the bit-slice will be obtained as the topdown Kronecker product of the transfer matrices of its gates. A particular severe constraint is the requirement that the controlling signals and the controlled gate must be in neighbour lines. This may require swapping some lines. Furthermore since the process is numerical, to evaluate the correct performance of a circuit with n inputs, all $2^n$ possible input vectors must be considered.

A special class of quantum computing circuits comprises circuits with dedicated target lines to do the actual processing and give the result of a computation, meanwhile all other lines do the controlling of the target gates. This kind of structure appears quite naturally when embedding a non-reversible function in a reversible one providing the target line(s). See examples below. To calculate the target output in this kind of circuits a low complexity symbolic calculus is introduced for circuits based on controlled-Q gates, which following the seminal work [5] are complex-valued matrices representing roots of NOT. Preliminary results of this work were presented at the 11th International Workshop on Boolean Problems [11].

*Formalisms:* Matrices giving formal representation to gates of a quantum computing circuit must be unitary, i.e. the product of a matrix and its adjoint must return the identity matrix. If the matrix is symmetric, the adjoint reduces to its complex conjugate. The matrix corresponding to the NOT operation is [0 1; 1 0], which is trivially unitary, since it is real, symmetric, and self-inverse. Therefore NOT* = NOT, leading to NOT·NOT* = NOT² = I, the identity matrix. For any k, power of 2, matrices Q, such that $Q^k$ = NOT, are also unitary and symmetric [1], [2]. Moreover, since NOT is selfinverse, $Q^{-k}$ = NOT$^{-1}$ = NOT and $Q^k = Q^{-k}$. Moreover, NOT·Q = NOT*·Q = $(Q^k)^*·Q$ = $(Q^k)^*·(Q^{-1})^*$ = $(Q^{k-1})^*$. It is easy to see that when k = 2, Q = V, and $(Q^{k-1})^* = Q^* = V^*$, as it is well known [1].

Notice that since Q is unitary and symmetric then, Q·Q* = I, from where it follows that Q* = $Q^{-1}$, Q = $(Q^{-1})^*$, Q = $(Q^*)^{-1}$ and $Q^0$ = I. Furthermore, since NOT = NOT* then: $Q^k = (Q^k)^* = (Q^{-1})^k = Q^{-k}$.

When k = 2, Q = V = $\frac{i+1}{2}\begin{bmatrix} 1 & -i \\ -i & 1 \end{bmatrix}$, and NOT·Q = $(Q^{k-1})^* = Q^* = V^*$.

When k = 4, Q = W = $\frac{1}{2}\begin{bmatrix} 1+i^{1/2} & 1-i^{1/2} \\ 1-i^{1/2} & 1+i^{1/2} \end{bmatrix}$, and NOT·Q = $(Q^{k-1})^* = (W^3)^*$

A special property that will be useful when evaluating circuits is the following, recalling that $Q^k$ = NOT and $Q^{\pm 2k}$ = I:

$$Q^{k(a \oplus b)c} = Q^{k(a+b)c} \cdot Q^{-2kabc} = Q^{k(a+b)c} \cdot I^{abc} = Q^{k(a+b)c}. \quad (1)$$

In the circuits shown below, an empty square will denote a gate which unitary matrix is Q, and a square with a diagonal line, a gate which unitary matrix is Q*. The circuits belong to the "CNQ" family, where "C" stands for "controlled", "N" for NOT and "Q", for the gates introduced earlier. It is easy to understand that CNV and CNW are special cases of CNQ.

Lemma: Given a two quantum bits –(qubits)- controlled-Q-gate with a binary control signal *a* and a target input *t*, then the target output *t'* equals $Q^a·t$.

Proof:
From the specification of a controlled gate, if *a* = 0 then the Q gate should be inhibited and behave as an identity, else, if *a* = 1 the Q gate is active and it will be applied to *t*.

This is exactly expressed with $t' = Q^a·t$, since if *a* = 0, then $Q^a = Q^0$ = I and therefore $t' = t$; meanwhile if *a* = 1, then $Q^a = Q^1$ = Q and $t' = Q·t$.

Remark: Since $\forall a_i \in \{0, 1\}$ with $1 \leq i \leq n$ holds:

$$Q^{a_1} \cdot Q^{a_2} \cdot ... \cdot Q^{a_n} = Q^{a_1+a_2+...+a_n} \quad (2)$$

and at the exponents level the sum is associative and commutative, this means that *for calculation purposes* the individual Q matrices may be reordered. Whether in the circuit a reordering of the CQs is possible will depend on the relative independence of the controlling expressions. Since the exponents of Q will be represented as GF(2) expressions possibly extended with Boolean operations, they will have to be translated to arithmetic expressions. Some basic elementary transformations are, *e.g.*: $1 \oplus a = 1 - a$; $a \oplus b = a + b - 2ab$ ; $a \oplus b \oplus ab = a + b - ab$; $a + b - (a \oplus b) = 2ab$.

*Examples*:

**Example 1**. Given a circuit with two cascaded Toffoli gates, as shown in Fig. 1, the question is, which is the lowest possible quantum cost for a realization based on CNV gates?

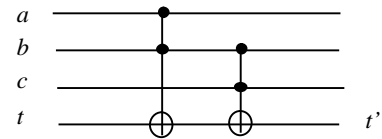

**Fig. 1** *Two cascaded Toffoli gates with a common target line*

From the functionality of the Toffoli gate it is simple to show that the target output will be $t' = t \oplus ab \oplus bc = t \oplus b(a \oplus c)$. A straightforward approach to obtain a CNV realization is to replace each Toffoli gate with the CNV circuit introduced in Barenco *et al.* [5], this leading to the circuit shown in Fig 2.

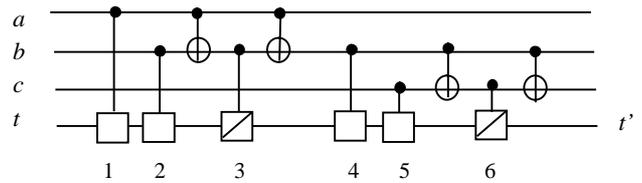

**Fig. 2** *CNV realization of two cascaded Toffoli gates*

A two-fold analysis of the realization will be done, to verify the functionality and to search for possible simplifications. (For this last purpose the CV and CV* gates have been numbered.). Since as mentioned in Lemma 1, the exponent of the controlled gates determine their behaviour, and considering the additivity of the exponents (Eq, (2)), the circuit analysis will be done at the level of exponents.

$t' = V^E·t$

$E = a + b - (a \oplus b) + b + c - (b \oplus c) = 2ab + 2bc = 2b(a + c)$

Therefore $t' = V^{2b(a+c)}·t$ and with Eq. (1), $t' = V^{2b(a \oplus c)}·t = $ NOT$^{b(a \oplus c)}·t$ $= t \oplus b(a \oplus c)$, as it was to be expected. The realization comprises 10 elementary controlled gates.

Note that the equation for E may also be given the following equivalent expression:

E = $a + (b + b) - (a \oplus b) + c - (b \oplus c)$, which represents movig the gate 4 towards the gate 2. That part of the circuit will have then a transfer of $V^b·V^b = V^{2b} = $ NOT$^b$. Hence, the two CV gates may be replaced by *one* CNOT gate, thus reducing the total number of controlled gates to 9, as shown in Fig. 3.

The former simple circuit transformation opens the question whether instead of a $(b + b)$ component, a $(b - b)$ were possible, which would lead to a cancellation of two gates. This would however imply that the gate 4 of Fig. 2, should have to be a CV* gate, and this would alter the behaviour of the circuit. It should however be recalled that since V*·V* also equals NOT, a Barenco *et al.* type of realization for a Toffoli gate may be done using CV* gates in the two first places and a CV gate at the third position (see e.g. [6]).

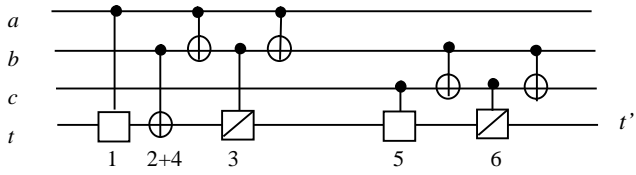

**Fig. 3** *Realization with 9 elementary controlled gates*

Fig. 4 shows the circuit with the above change for the second Toffoli gate.

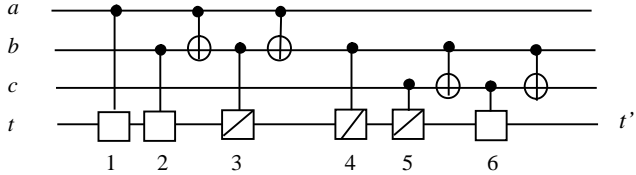

**Fig. 4** *Realization that allows cancelling gates 2 and 4*

$t' = V^E \cdot t$

$E = a + b - (a \oplus b) - b - c + (b \oplus c) = 2ab - 2bc = 2b(a-c)$
$= a + (b-b) - (a \oplus b) - c + (b \oplus c)$
$= a - (a \oplus b) - c + (b \oplus c)$ (after cancelling gates 2 and 4)
$= 2b(a-c)$

$t' = V^{2b(a-c)} = NOT^{b(a-c)} = NOT^{ab} \cdot NOT^{-bc} = NOT^{ab} \cdot NOT^{bc} =$
$= NOT^{b(a+c)} = NOT^{b(a \oplus c)}$   // Recall Eq. (1)
$= t \oplus b(a \oplus c)$

The functionallity is correct and the realization requires 8 gates.

However, we *claim* that the circuit of Fig. 5 realizes the two Toffoli gates in cascade, but requiring only 7 gates.

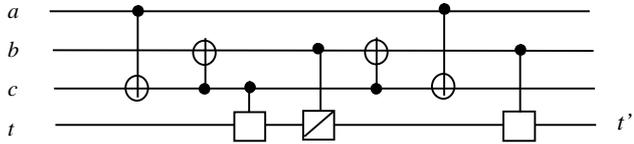

**Fig. 5** *Realization of the two cascaded Toffoli gates with only 7 controlled gates*

Proof:
$t' = V^E \cdot t$
$E = (a \oplus c) - (a \oplus b \oplus c) + b = (a \oplus c) - (a \oplus c) - b + 2b(a \oplus c) + b$
$= 2b(a \oplus c)$
$t' = V^{2b(a \oplus c)} \cdot t = NOT^{b(a \oplus c)} \cdot t = t \oplus b(a \oplus c)$       q.e.d.

**Example 5**. (Adapted from [4])

The circuit of Fig. 6 should be a realization of two Toffoli gates with two and three controls and different not independent target lines. The realization is based on the so called CNVW library, i.e. both control-V and control-W gates are used. (In Fig. 6, CW gates are represented with solid lines, the CV gates, with dash-lines). Note that the target lines are not totally independent. Since $W^2 = V$, the symbolic verification will be expressed in terms of W and $W^2$ gates. Working at the level of exponents, this means that the exponents of V gates have to be simply multiplied by 2 to represent their functionality in terms of W gates.

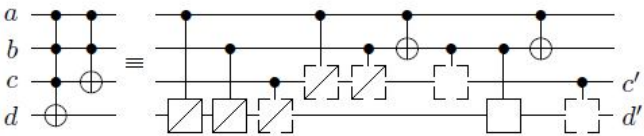

**Fig. 6** *CNVW circuit to be evaluated.*

$c' = W^E \cdot c$
$E = -2a - 2b + 2(a \oplus b) = -2(a + b - (a \oplus b)) = -4ab$
$c' = W^{-4ab} \cdot c = NOT^{ab} \cdot c = c \oplus ab$

The output of the Toffoli gate with two control signals is correct. Note that this signal controls the last gate on the other target line.

$d' = W^{E'} \cdot d$
$E' = -a - b - 2c + (a \oplus b) + 2c' = -a - b - 2c + (a \oplus b) + 2(c \oplus ab)$
$= -a - b - 2c + (a \oplus b) + 2(c + ab - 2abc) =$
$= -a - b + (a \oplus b) + 2ab - 4abc = -2ab + 2ab - 4abc = -4abc$
$d' = W^{-4abc} \cdot d = NOT^{abc} \cdot d = d \oplus abc$

The output of the Toffoli gate with three control signals is also correct. It should be pointed out that the realization of Fig. 6 requires only 10 elementary controlled gates. If the control signal *c* should be recovered, an additional CNV Toffoli gate would be needed to add *ab* to *c'* in mod 2, and recover *c*. The elements of the additional Toffoli gate may be reordered leaving the CNOT gate with target line *b* in the first position, thus cancelling the CNOT gate on line *b* at the end of the original circuit. This would lead to a realization with 13 elementary gates, which is optimal, according to the classical CNW realization [5].

*Conclusion:*

A symbolic calculus has been introduced, which uses powers of the unitary matrices of the CNQ library. The exponents of the matrices are Boolean expressions. It was shown that this is an appropriate formalism for the otherwise verbal especification of the behavior of a controlled gate. The calculus was developed for the evaluation of the performance of target line oriented reversible/quantum circuits based on the family of controlled Q gates. Although the considered circuits involve complex-valued matrices the calculus only uses Boolean expressions and plain arithmetic. The proposed calculus is much simpler that the one presented in [7], and also much simpler than evaluations based on bit-slices.

*Acknowledgments:* The work of C. Moraga was partially supported by the CICYT Spain, under project TIN 2011-29827-C02-01.

F.Z. Hadjam (*European Centre for Soft Computing*. 33600 Mieres, Spain)

C. Moraga (*European Centre for Soft Computing*. 33600 Mieres, Spain)
E-Mail: claudio.moraga@softcomputing.es

F.Z. Hadjam: also with *Département Informatique*, *Djillali Liabes University*, Sidi Bel Abbes, Algeria

C. Moraga: also with *Faculty of Computer Science, TU Dortmund University*, Dortmund, Germany.